%% file: sample-sigconf.tex
\documentclass[sigconf,acmthm=false]{acmart}

\usepackage{latexsym}
\usepackage[T1]{fontenc}
\usepackage[utf8]{inputenc}
\usepackage{microtype}
\usepackage{graphicx}

\usepackage{url}
\usepackage{xcolor}
\usepackage{wrapfig}
\usepackage{enumitem}
\usepackage{multirow}
\usepackage{natbib}
\usepackage{booktabs}
\usepackage{arydshln}
\usepackage{appendix}
\usepackage{array}
\usepackage{stfloats}
\usepackage[table]{xcolor}
\usepackage{longtable}
\usepackage{mdframed}
\usepackage{ltablex} 
\keepXColumns
\usepackage{caption}
\usepackage{makecell} 
\usepackage{bm}
\usepackage{ragged2e}  
\usepackage{thm-restate}
\makeatletter
\@ifundefined{newcounteralias}{}{%
  \renewcommand\thmt@autorefsetup{\@xa\def\csname\thmt@envname autorefname\@xa\endcsname\@xa{\thmt@thmname}}%
}
\makeatother

\usepackage{algorithm}
\usepackage{algpseudocode}

\theoremstyle{plain}
\newtheorem{theorem}{Theorem}
\newtheorem{proposition}[theorem]{Proposition}

\definecolor{myblue}{RGB}{235, 244, 246}
\definecolor{mygreen}{RGB}{234, 240, 225}

\usepackage[most]{tcolorbox}

\newcounter{myboxcounter}
\renewcommand{\themyboxcounter}{\arabic{myboxcounter}}
\newtcolorbox{custombox_red}[2][]{
    colback=red!5!white,
    colframe=red!75!black,
    fonttitle=\bfseries,
    title=Box~\themyboxcounter: #2,
    enhanced,
    #1
}

\newtcolorbox{custombox_orange}[2][]{
    colback=orange!5!white,
    colframe=orange!75!black,
    fonttitle=\bfseries,
    title=Box~\themyboxcounter: #2,
    #1
}
\newtcolorbox{custombox_blue}[2][]{
    colback=blue!5!white,
    colframe=blue!75!black,
    fonttitle=\bfseries,
    title=Box~\themyboxcounter: #2,
    #1
}
\newtcolorbox{custombox_green}[2][]{
    colback=green!5!white,
    colframe=green!75!black,
    fonttitle=\bfseries,
    title=Box~\themyboxcounter: #2,
    #1
}
\newtcolorbox{custombox_black}[2][]{
    colback=black!5!white,
    colframe=black!75!black,
    fonttitle=\bfseries,
    title=Box~\themyboxcounter: #2,
    #1
}

\usepackage[most]{tcolorbox}

\newcommand{\eg}{\emph{e.g., }}

\newcommand{\cf}{\emph{cf. }}

\definecolor{ForestGreen}{RGB}{34,139,34}
\renewcommand{\citet}[1]{\citeauthor{#1} [\citeyear{#1}]}

\newcolumntype{C}[1]{>{\centering\arraybackslash}p{#1}}

\setcopyright{acmlicensed}
\copyrightyear{2027}
\acmYear{2027}
\acmDOI{XXXXXXX.XXXXXXX}
\acmConference[Conference acronym 'XX]{Make sure to enter the correct conference title from your rights confirmation email}{June 03--05,
  2018}{Woodstock, NY}

\begin{document}
\begin{sloppypar}

\title{EchoRec: Multi-Item Prediction-Empowered Generative Recommendation via Cycle-Consistent Preference Alignment}

\author{Haokai Ma}
\email{haokai.ma1997@gmail.com}
\authornotemark[1]
\thanks{*Equal Contribution}
\affiliation{%
 \institution{National University of Singapore}
 \city{Singapore}
 \country{Singapore}}

\author{Aoqi Hu}
\email{huaoqi800@gmail.com}
\authornotemark[1]
\affiliation{%
 \institution{National University of Singapore}
 \city{Singapore}
 \country{Singapore}}

\author{Yueao Xing}
\email{sd513@bupt.edu.cn}
\affiliation{%
 \institution{Beijing University of Posts and Telecommunications}
 \city{Beijing}
 \country{China}}

\author{Ruobing Xie}
\email{xrbsnowing@163.com}
\affiliation{%
 \institution{Tencent}
 \city{Beijing}
 \country{China}}

\author{Yonghui Yang}
\email{yyh.hfut@gmail.com}
\affiliation{%
  \institution{National University of Singapore}
  \city{Singapore}
  \country{Singapore}}

\author{Teng Tu}
\email{teng.tu@u.nus.edu}
\affiliation{%
 \institution{National University of Singapore}
 \city{Singapore}
 \country{Singapore}}

\author{Lei Meng}
\email{lmeng@sdu.edu.cn}
\affiliation{%
 \institution{Shandong University}
 \city{Jinan}
 \country{China}}

\author{Tat-Seng Chua}
\email{dcscts@nus.edu.sg}
\affiliation{%
 \institution{National University of Singapore}
 \city{Singapore}
 \country{Singapore}}
 
\renewcommand{\shortauthors}{Haokai Ma et al.}

\begin{abstract}
Generative recommendation autoregressively generates the semantic IDs of the target item, unifying preference modeling and index retrieval within the shared token space. Recent attempts have introduced Multi-Token Prediction (MTP) into this field, yet they primarily inherit its efficiency merit, leaving its potential as dense supervision unexplored. Unlocking this potential hinges on whether future behaviors qualify as informative supervision. Our analysis reveals that future behaviors carry a semantic echo of the current one far above that of random pairs, which nevertheless decays along horizons under intent transitions, making them informative yet order-dependent signals. Motivated by this, we propose \textbf{EchoRec}, which \textbf{e}mpowers MTP with \textbf{c}ycle-consistent \textbf{h}olistic preference alignment across multi-h\textbf{o}rizon for generative \textbf{rec}ommendation. It comprises two synergistic modules. Horizon-aware Preference Generation (HPG) sequentially chains lightweight auxiliary branches upon the base recommender, where each branch conditions on its predecessor to respect preference evolution. Verifiable Holistic-Preference Alignment (VHA) further consolidates them into the holistic preference and echoes it back through cycle-consistent projectors to suppress spurious alignment, with a theoretical guarantee that excludes the rank-collapse form of spurious alignment under an invertible round trip, enabling the holistic preference to be retained in the decoding representation. All auxiliary components serve as disposable scaffolding discarded at inference, introducing negligible online serving overhead. Extensive experiments on four datasets demonstrate the superiority of our \textbf{EchoRec}, together with its naturally acquired multi-item generation ability. Our code and datasets will be available upon acceptance.
\end{abstract}

\ccsdesc[500]{Information systems~Recommender systems}
\keywords{Multi-Token Prediction, Generative Recommendation} 



\maketitle

\begingroup
\let\oldaddcontentsline\addcontentsline
\renewcommand{\addcontentsline}[3]{}
\input{Sections/1_Introduction}
\input{Sections/2_Related_Works}
\input{Sections/3_Preliminary}
\input{Sections/4_Methodology}
\input{Sections/5_Experiments}
\input{Sections/6_Conclusion}
\endgroup

\balance
\bibliographystyle{ACM-Reference-Format}
\bibliography{sample-base}

\end{sloppypar}
\end{document}

%% file: Sections/1_Introduction.tex
\section{Introduction}

Generative recommendation (GR) has recently emerged as a promising paradigm that tokenizes each item into a tuple of discrete semantic IDs (SIDs) and generates the target item token by token~\cite{GR1,GR8}. Instead of ranking candidates over the entire corpus, GR unifies preference modeling and index retrieval within a single process, where the shared token space anchors semantically related items to overlapping SIDs~\cite{GR3,GR7}. It enables recommenders to function well in sparse scenarios by exploiting external semantics, attracting increasing attention from e-commerce and streaming services. 

Drilling down into this topic, existing works~\cite{GR6} build the GR pipeline around item tokenization and autoregressive generation. Regarding the former, TIGER~\cite{TIGER} generates item SIDs via Residual Quantized Variational Autoencoder (RQ-VAE), LETTER~\cite{LETTER} regularizes the tokenizer with collaborative signals and code assignment diversity, and Pctx~\cite{Pctx} integrates user contexts for personalization. For the latter, EAGER~\cite{EAGER} integrates behavioral and semantic information through two-stream decoding with contrastive objectives. ETEGRec~\cite{ETEGRec} unifies both components via alternating optimization with sequence-item and preference-semantic alignment.

\begin{figure}[!t]
    \includegraphics[width=\linewidth]{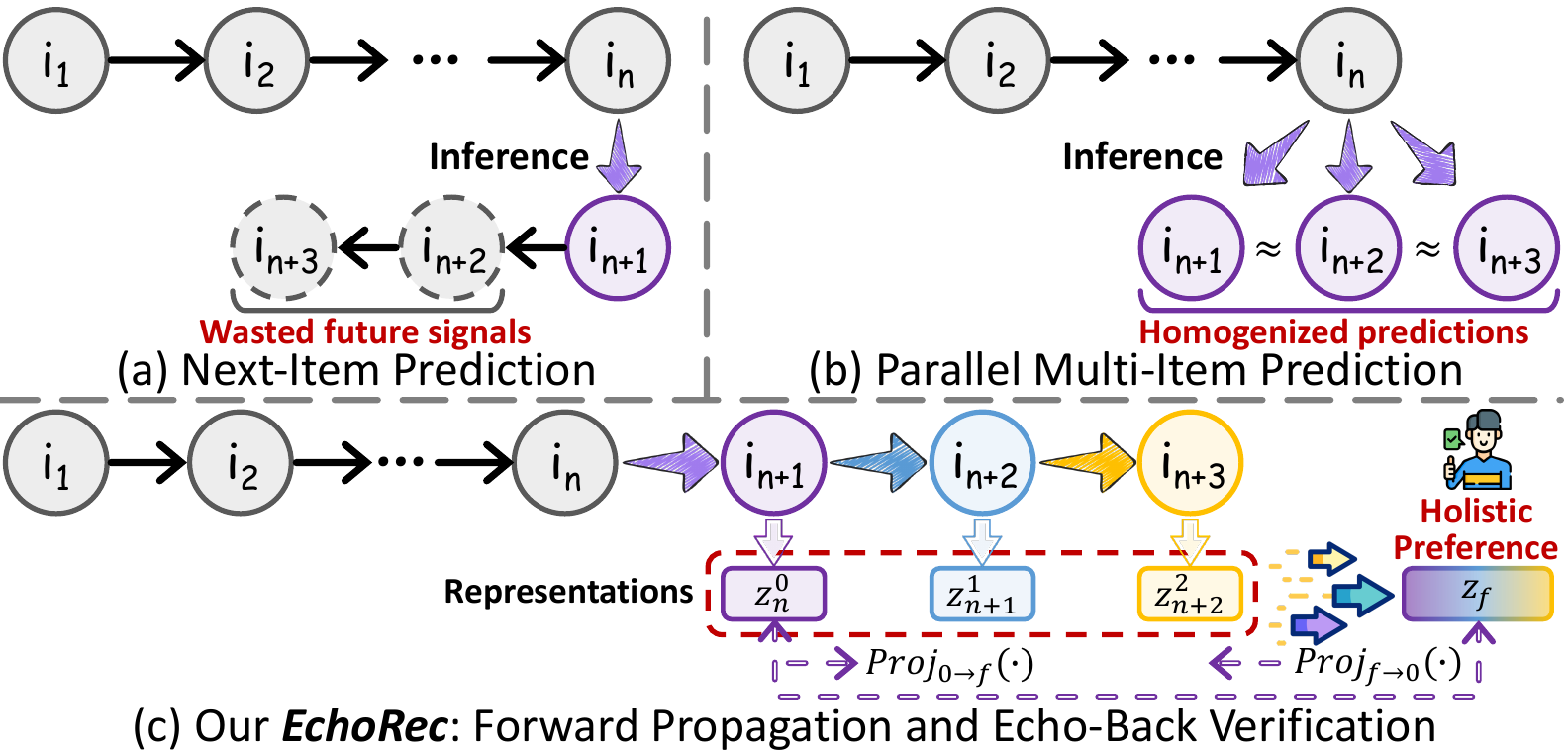}
    \vspace{-0.4cm}
    \caption{Conceptual comparison of future-behavior modeling paradigms. (a) Next-item prediction underutilizes future-behavior supervision. (b) Parallel multi-item prediction risks homogenizing the horizons. (c) \textbf{EchoRec} propagates supervision across horizons and cycle-consistently echoes the decoding representation with holistic preference.}
    \vspace{-0.6cm}
    \label{fig:motivation_graph}
\end{figure}

Inspired by the remarkable capabilities of Multi-Token Prediction (MTP) in advanced Large Language Models (LLMs)~\cite{Meta-MTP,Deepseek-MTP}, several pioneering efforts have introduced it into recommendation. For instance, GReF~\cite{GReF} generates multiple ordered future items with MTP for reranking efficiency, RPG~\cite{RPG} applies MTP to expand the token number of items to boost expressiveness, and GPR~\cite{GPR} and EGA-V2~\cite{EGA-V2} adopt MTP to capture concurrent interests and generate advertisement content in industrial scenarios. Notably, beyond the decoding efficiency that MTP is best known for, it also serves as dense supervision that densifies training signals and encourages pre-planned representations~\cite{pitfalls,MTP1}. However, existing efforts primarily transplant the former merit, leaving the latter potential as future-behavior supervision largely unexplored.

However, translating such potential into GR faces two fundamental challenges: (1) \textit{\textbf{How to exploit future behaviors as sequentially dependent supervision across horizons?}} Unlike consecutive tokens coupled by grammar in LLMs, consecutive behaviors are driven by evolving user intents, where $i_{n+2}$ depends on the intent transition triggered by $i_{n+1}$ rather than on the history alone. Several ID-based recommenders that exploit future interactions~\cite{GRec,Oracle4Rec,FENRec} merely treat them as auxiliary signals of next-item prediction. Meanwhile, MTP-empowered recommenders either repurpose MTP as an efficiency-oriented device~\cite{RPG,GPR,EGA-V2} or decode multiple items simultaneously from the same context~\cite{GReF}, degenerating multi-item prediction into homogenized next-item prediction that disregards preference evolution. (2) \textit{\textbf{How to verify that the future signals are actually internalized?}} Even equipped with future-aware supervision, such signals shape the decoding representation only indirectly through auxiliary objectives. A natural remedy is to align it toward the aggregated holistic target via a projector, yet such one-way alignment is inherently \emph{unverifiable}, as the projector can absorb the subspace discrepancy to satisfy it while leaving the representation oblivious to the holistic preference, a failure mode that we term \textbf{spurious alignment}. To our knowledge, no existing GR study offers a mechanism to guard against it.

To address these challenges, we propose \textbf{EchoRec} that \textbf{e}mpowers MTP with \textbf{c}ycle-consistent \textbf{h}olistic preference alignment across multi-h\textbf{o}rizon for GR. Drawing an analogy to the acoustic echo, \textbf{EchoRec} regards preference modeling as an echoing process: user preferences are propagated forward across future horizons as sequentially dependent supervision and echoed back through a cycle-consistent round trip to verify their internalization. Specifically, we devise the \emph{Horizon-aware Preference Generation} (HPG) module to extend the next-item objective to multiple future horizons through sequential MTP branches, each of which conditions on its predecessor's representation to capture intent transition between adjacent horizons, thereby delivering the structural-level supervision that respects preference evolution. To further promote such internalization, we design a \emph{Verifiable Holistic-Preference Alignment (VHA)} module, which consolidates all branches into the holistic preference and pulls the decoding representation toward it. Considering that such one-way alignment remains unverifiable, VHA further imposes a cycle-consistency constraint on the projector pair, whose checkable invertibility provably excludes the rank-collapse form of spurious alignment (\cf Proposition~\ref{proof:pro}). Notably, EchoRec retains only the auxiliary branches of HPG for multi-item generation, which reuse the shared output head and decoding graph. For next-item prediction, these branches are further discarded, leaving its serving cost nearly identical to that of the base recommender.

We conduct extensive experiments on four real-world datasets to demonstrate the effectiveness of our \textbf{EchoRec}. We further conduct analyses on multi-horizon prediction, component ablation, and efficiency to verify its robustness, universality and the superiority of each design. The main contributions are as follows:
\begin{itemize}[leftmargin=*, topsep=0.2pt,parsep=0pt]
    \item We propose \textbf{EchoRec} to empower MTP with cycle-consistent holistic preference alignment to capture horizon-aware preferences. To our knowledge, it is among the pioneers to unlock MTP as sequentially dependent future-behavior supervision in GR.
    \item We devise HPG to inject sequentially dependent supervision across horizons and VHA to verifiably internalize the holistic preference via cycle-consistent projectors, whose checkable invertibility provably excludes rank-collapse spurious alignment.
    \item We conduct extensive experiments on four datasets, where \textbf{EchoRec} consistently improves diverse SOTA backbones while exhibiting the multi-item generation ability.
\end{itemize}

%% file: Sections/2_Related_Works.tex
\section{Related Works}
\noindent
\textbf{Generative Recommendation.} 
Inspired by the remarkable achievement of generative models, pioneering recommenders assign each item a set of discrete yet shared tokens, termed \emph{semantic ID} (SID), where tokenization discretizes item semantics into SIDs and generation predicts the target SIDs. Specifically, VQ-Rec~\cite{VQ-Rec} and SeeDRec~\cite{SeeDRec} respectively represent items with transferable ``codes'' and ``sememes'' from their content. TIGER~\cite{TIGER} derives SIDs via RQ-VAE and predicts the target item with a Transformer-based generator. EAGER~\cite{EAGER} and LETTER~\cite{LETTER} align behavioral and semantic information via contrastive objectives. DiscRec~\cite{DiscRec} injects position embeddings and disentangles semantic and collaborative signals, while Pctx~\cite{Pctx} devises a context-aware tokenizer for personalized SIDs. ETEGRec~\cite{ETEGRec} further unifies both stages via alternating optimization with sequence-item and preference-semantic alignment. However, these efforts generally follow the next-item objective, leaving future behaviors beyond the immediate target untouched.

\noindent
\textbf{Future-Aware Recommendation.}
As future interactions reveal preference evolution, several sequential recommenders incorporate them into preference learning. GRec~\cite{GRec} adopts a gap-filling encoder-decoder that encodes bidirectional contexts to predict masked items. DualRec~\cite{DualRec} mutually aligns past and future encoders, while Oracle4Rec~\cite{Oracle4Rec} enables the future encoder to guide the past one. FENRec~\cite{FENRec} converts subsequent items into position-decayed soft labels of the next-item objective. However, these ID-based methods treat future interactions as auxiliary signals of next-item prediction, overlooking the intent transitions across horizons.

\noindent
\textbf{Multi-Token Prediction for Recommendation.}
Multi-Token Prediction (MTP) extends the next-token objective of LLMs to predict multiple tokens per step, alleviating their inefficiency and limited long-range dependency modeling~\cite{MTP1,MTP2,MTP3,MTP4}. Inspired by its superiority, several attempts have integrated MTP into recommendation. GReF~\cite{GReF} generates multiple ordered future items simultaneously for real-time reranking efficiency. RPG~\cite{RPG} allocates longer SIDs to each item and decodes them in parallel to enhance item expressiveness. GPR~\cite{GPR} unifies advertisements and organic content in a shared SID space and integrates MTP to capture concurrent interests, while EGA-V2~\cite{EGA-V2} chains MTP branches across modalities to jointly generate advertisement sequences and creatives. Nevertheless, these works predominantly inherit the efficiency merit of MTP, whereas our work unlocks its potential as sequentially dependent future-behavior supervision with verifiable internalization.

%% file: Sections/3_Preliminary.tex
\section{Preliminary}
\label{sec.preliminary}

\subsection{Multi-Token Prediction}
As the name implies, Multi-token prediction (MTP) extends the conventional next-token prediction by predicting multiple future tokens within a single step, thereby providing longer-range dependencies during training. Given the token sequence $x_{1:t}$, the generic MTP objective can be written as $\mathcal{L}_{\mathrm{MTP}}\!=\!-\sum_{t} \log P_{\theta}(x_{t+1:t+D}\!\mid\! x_{1:t})$, where $D$ denotes the prediction depth. A representative implementation~\cite{Meta-MTP} adopts a shared Transformer trunk $f_s(\cdot)$ with $D$ independent prediction heads $f_h^{(i)}(\cdot)$ to predict these future tokens in parallel from the same contextual representation. Let $z_{1:t}\!=\!f_s(x_{1:t})$ represent the hidden state produced by $f_s(\cdot)$, the $i$-th future token can be predicted as $P_{\theta}(x_{t+i}\!\mid\! x_{1:t})\!=\!\mathrm{softmax}\big(f_u(f_h^{(i)}(z_{1:t}))\big)$, where $f_u(\cdot)$ denotes the shared unembedding matrix. Instead of predicting all future tokens independently from the same state, DeepSeek-V3~\cite{Deepseek-MTP} sequentially chains MTP modules to preserve the causal dependencies among future predictions and sequentially predict additional $D$ tokens. To be specific, the $b$-th MTP module combines its predecessor's representation with the embedding of the $(i\!+\!b)$-th token to generate the input $\bm{h}_i^{\prime b}$ of the additional Transformer block:
\begin{equation}
\label{eq:mtp_deepseek}
\bm{h}_i^{\prime b}=M_b\!\left[\mathrm{RMSNorm}(\bm{h}_i^{b-1});\mathrm{RMSNorm}(\mathrm{Emb}(i_{i+b}))\right],
\end{equation}
where $\mathrm{RMSNorm}(\cdot)$, $M_b(\cdot)$, and $\mathrm{Emb}(\cdot)$ denote the Root Mean Square Layer Normalization, the projection matrix and the shared embedding layer, respectively. The Transformer block $\mathrm{TRM}_b(\cdot)$ then produces the output representation $\bm{h}_{i+b}^{b}$, upon which the shared output head predicts the $(i\!+\!b\!+\!1)$-th token. We adopt this sequential design to preserve coherent causal dependencies across multiple horizons, providing denser supervision over future tokens while keeping the auxiliary objective compatible with generation.

\begin{figure}[!t]  
  \centering
  \includegraphics[width=\columnwidth]{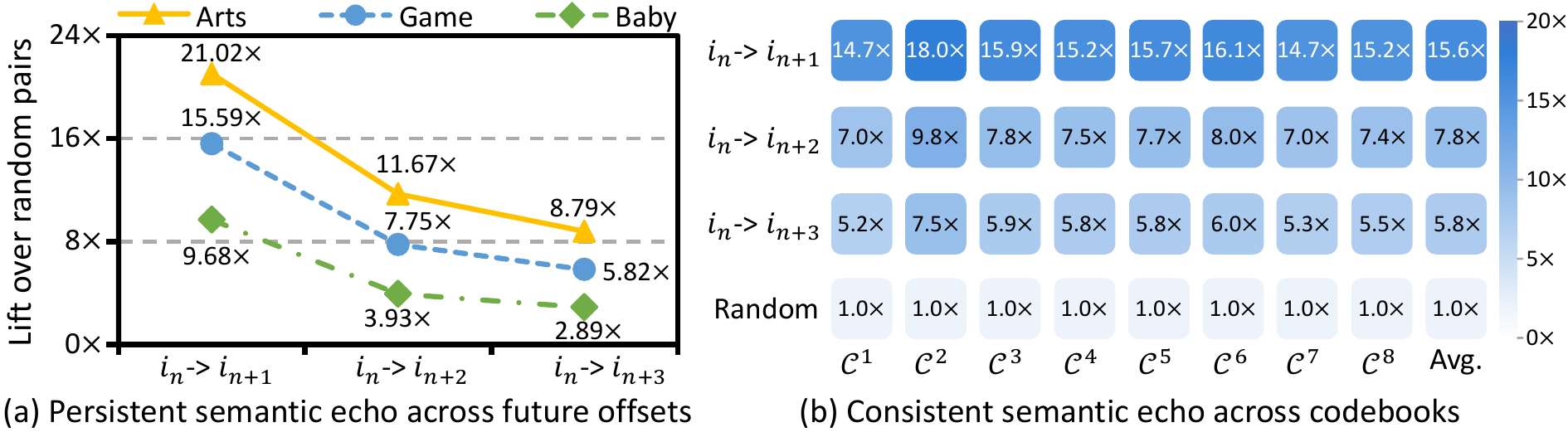}
  \vspace{-0.5cm}
  \caption{Informativeness analysis of future behaviors for next-item prediction. (a) SID overlap lift of $(i_n,i_{n+h})$ over random pairs on three datasets, remaining well above random even at larger offsets. (b) Codebook-wise lift on Game, where the semantic echo stays consistent across all codebooks.}
\label{fig:necessary}
\vspace{-0.6cm}
\end{figure}

\subsection{Motivation Analysis on Future Behaviors}
\label{subsec:preliminary}
Exploiting future behaviors as auxiliary supervision is appealing yet non-trivial, which hinges on two fundamental questions: \textit{(1) whether future behaviors convey informative signals for next-item prediction, rather than mere noise induced by drifting intents} and \textit{(2) whether multiple future behaviors can be faithfully forecast in parallel from the shared context, rather than collapsing into homogenized variants?} 
We conduct two motivation analyses to respectively answer them.

\subsubsection{Semantic Echo across Future Behaviors}
\label{subsubsec:preliminary_1}
To answer the first question, we measure the SID overlap of each observed pair $(i_n, i_{n+h})$ with offset $h\!\in\!\{1,2,3\}$ on three datasets, and illustrate the \textit{lift} in Figure~\ref{fig:necessary}, which is defined as the ratio of this overlap to that of randomly sampled item pairs. The \textit{lift} of $1.0\times$ indicates that the future item $i_{n+h}$ is semantically indistinguishable from a random one, whereas a higher \textit{lift} indicates the future behaviors are semantically anchored to the current interaction $i_n$. As shown in Figure~\ref{fig:necessary} (a), \textbf{future behaviors $i_{n+h}$ persistently echo the related semantics of the current interaction $i_n$ across all offsets}: the \textit{lift} reaches up to $21.02\times$ at $h{=}1$ and remains $2.89\times$--$8.79\times$ even at $h{=}3$ across three datasets. This indicates that interactions beyond $i_{n+1}$ still echo the semantics of the current one rather than degenerating into noise, thus potentially constituting informative supervision for preference modeling. Besides, the consistent decay along $h$ reveals that this echo is progressively reshaped by intent transitions, suggesting that different horizons carry order-dependent signals and should be modeled as sequentially dependent targets rather than being treated equally. Moreover, Figure~\ref{fig:necessary} (b) shows that \textbf{such semantic echo remains consistent across all codebooks}, where all eight codebooks exhibit comparable \textit{lift} on Game across all offsets, indicating that such semantic echo is not driven by certain dominant codebooks but consistently spans all semantic subspaces.

\begin{figure}[!t]  
  \centering
  \includegraphics[width=\columnwidth]{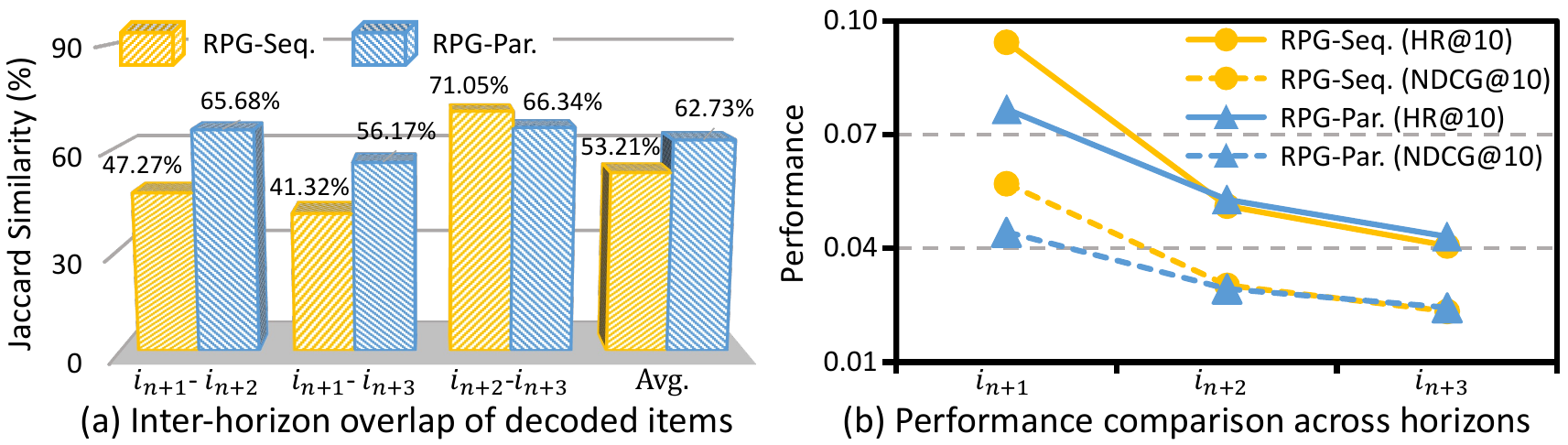}
  \vspace{-0.5cm}
  \caption{Analysis of the homogenization pitfall of parallel MTP structures on Game. (a) Inter-horizon Jaccard similarity of top-10 decoded items, where \emph{RPG-Par.} exhibits higher overlap than \emph{RPG-Seq.} on average. (b) Performance across horizons, where \emph{RPG-Par.} merely matches \emph{RPG-Seq.} at farther horizons yet underperforms at the immediate horizon $i_{n+1}$.}
\label{fig:Sequential_vs_Parallel}
\vspace{-0.6cm}
\end{figure}

\subsubsection{Homogenization Pitfall of Parallel MTP Structures}
\label{subsubsec:preliminary_2}
To answer the second question, we instantiate two multi-horizon variants upon RPG~\cite{RPG}, where \emph{RPG-Par.} decodes three future items from the shared context following the parallel MTP structure~\cite{Meta-MTP}, while \emph{RPG-Seq.} autoregressively rolls out the prediction as the input sequence to decode the next horizon. Figure~\ref{fig:Sequential_vs_Parallel}~(a) reports the Jaccard similarity between the top-10 predictions decoded at different horizons on Game, showing that \textbf{parallel MTP homogenizes multi-horizon predictions}: \emph{RPG-Par.} yields higher average inter-horizon overlap than \emph{RPG-Seq.} (62.73\% vs. 53.21\%), especially on the pairs involving the immediate horizon $i_{n+1}$ (65.68\% vs. 47.27\% for $i_{n+1}$--$i_{n+2}$ pair and 56.17\% vs. 41.32\% for $i_{n+1}$--$i_{n+3}$ pair). This indicates that predicting all horizons from the same context blurs the intermediate intent transition, whereas even the naive rollout that conditions each horizon on its predecessor differentiates the predictions across horizons better. Figure~\ref{fig:Sequential_vs_Parallel} (b) further compares the performance of these two variants at each horizon, revealing that \textbf{such homogenization fails to translate into performance gains}. Despite the dedicated supervision on each horizon, \emph{RPG-Par.} merely matches the unsupervised rollout of \emph{RPG-Seq.} at $i_{n+2}$ and $i_{n+3}$, yet underperforms at the immediate horizon $i_{n+1}$ that directly determines the recommendation quality. This clearly indicates that the future signals injected by parallel MTP are not effectively absorbed into horizon-specific predictions, and instead interfere with the immediate objective. 

%% file: Sections/4_Methodology.tex
\section{Methodology}
\label{sec:methodology}
\begin{figure*}[!t]
    \includegraphics[width=0.9\linewidth]{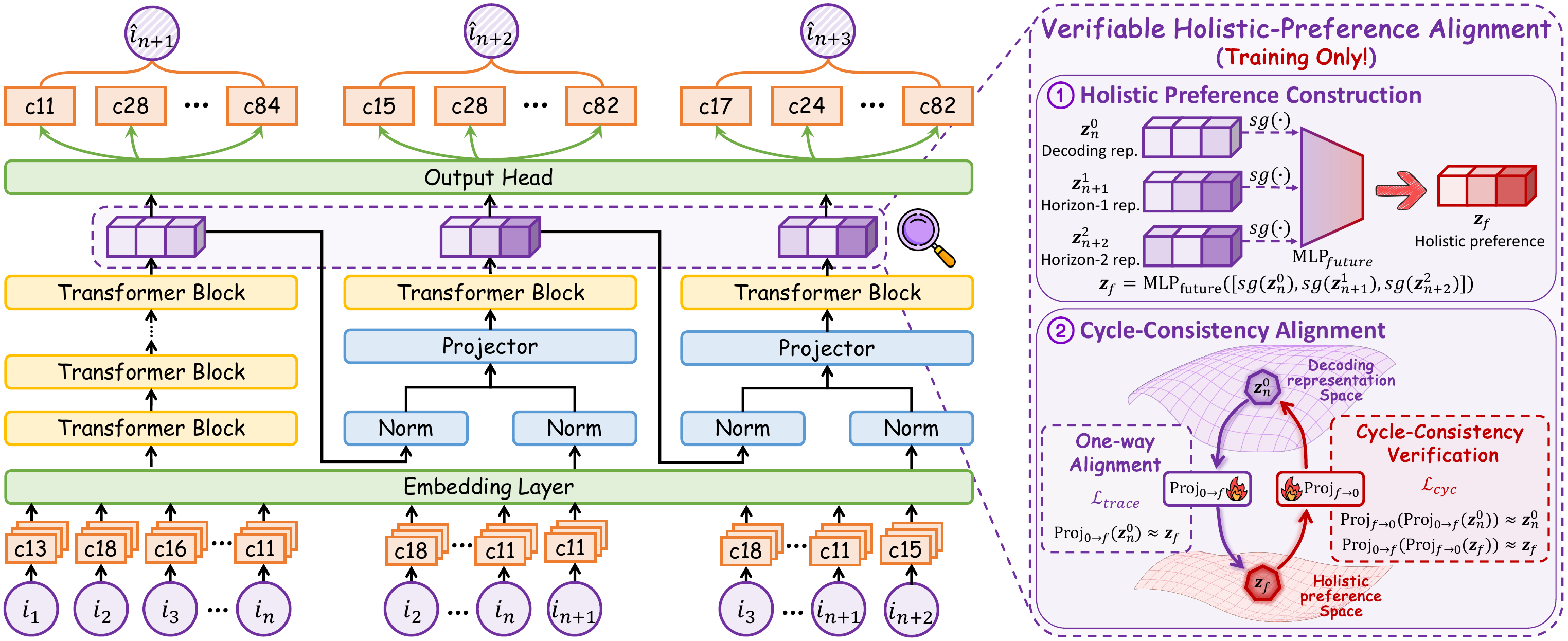}
    \vspace{-0.3cm}
    \caption{Overall structure of our \textbf{EchoRec}.}
    \vspace{-0.5cm}
    \label{fig:Overall_Structure}
\end{figure*}

\subsection{Problem Formulation}
\label{sec:problem_formulation}
We define the \emph{behavioral sequence} of user $u\!\in\!\mathcal{U}$ as $S_u\!=\!\{i^u_1,\cdots,i^u_n\}$, where $i^u_t\!\in\!\mathcal{V}$ denotes the $t$-th behavior of $S_u$ and $n$ denotes the sequence length. We adopt optimized product quantization~\cite{ge2013optimized, VQ-Rec} to represent item $i_t$ as a $k$-token tuple $\phi(i_t)$$\!=\!(c_{(t,1)},c_{(t,2)},\cdots,c_{(t,k)})$, where $\phi(i_t)$ denotes the SID of item $i_t$ and $c_{(t,j)} \!\in\! \mathcal{C}^j$ denotes the code selected from the $j$-th codebook $\mathcal{C}^j$. Given $S_u$ and the SIDs of all involved items, our goal is to predict the $k$-token tuple $\phi(i^u_{n+1})$ of the target item $i_{n+1}$\footnote{For brevity, we omit the superscript $u$ in $i^u_t$ in the following sections.} that user $u$ will interact with.\footnote{The key notations are summarized in Table~\ref{tab:notation} of Appendix.}

\subsection{Overall Structure}
\label{sec:overview}
The structure of \textbf{EchoRec} is illustrated in Figure~\ref{fig:Overall_Structure}, which regards preference modeling as an acoustic echoing process. Specifically, HPG (\cf Section~\ref{sec:HPG}) propagates user preferences forward across future horizons, extending the next-item objective through sequential MTP branches where each branch conditions on its predecessor to capture the intent transition. VHA (\cf Section~\ref{sec:VHA}) then echoes them back for verification, pulling the decoding representation toward the consolidated holistic preference under a cycle-consistency constraint that suppresses spurious alignment.

\subsection{Base Recommender}
\label{sec:base_recommender}
We adopt RPG~\cite{RPG} as the base recommender to tokenize each item using a long semantic ID and generate its tokens in parallel. Following Section~\ref{sec:problem_formulation}, each item $i_t$ is encoded as $\phi(i_t)=(c_{(t,1)},\cdots,c_{(t,k)})$, with $c_{(t,j)} \in \mathcal{C}^j$. To avoid inflating the context length under such long SIDs, we aggregate all token embeddings of item $i_t$ into a single representation $\bm{v}_{i_t}\!=\!\mathrm{Agg}(\bm{e}_{(t,1)},\cdots,\bm{e}_{(t,k)})$ via pooling. Given a user sequence $S_u$, the corresponding item representations $\bm{H}\!=\![\bm{v}_1,\bm{v}_2,\cdots,\bm{v}_n]$ are encoded by multiple Transformer blocks into the sequence representation $\bm{s}$ at position $n$. The shared output head $g(\cdot)$ then splits its output into $k$ codebook-specific vectors, and the token-level distribution of $j$-th codebook is computed as
\begin{equation}
\label{eq:token_prob}
P_{j}=\operatorname{softmax}\big(\bm{E}^{j}\cdot[g(\bm{s})]_{j}/\tau\big),\quad j=1,\cdots,k,
\end{equation}
where $[g(\bm{s})]_{j}$ denotes the $j$-th split vector, $\bm{E}^{j}$ is the embedding matrix of $\mathcal{C}^j$, and $\tau$ is the temperature. Assuming the tokens of target item $i_t$ to be conditionally independent given $\bm{s}$, we optimize the token-level MTP loss $\mathcal{L}_{\mathrm{RPG}}\!=\!-\sum_{j=1}^{k}\log P_j(c_{(t,j)})$ to score the target SID against all token combinations across codebooks~\cite{RPG}. For inference, the graph-constrained decoding samples an initial beam of valid SIDs from $\{P_{j}\}_{j=1}^{k}$, propagates it over a similarity graph $\mathcal{G}$, scores each candidate by $\sum_{j=1}^{k}\log P_{j}(c_j)$, and retains the top-$B$ candidates per step, yielding the top-$K$ recommendations without exhaustive scoring over the entire item pool. 

\subsection{Horizon-aware Preference Generation}
\label{sec:HPG}
Although existing MTP-based recommenders improve decoding efficiency or item expressiveness, most of them still follow the next-item prediction paradigm. However, user behaviors are inherently sequential and evolving, where the later ones beyond $i_{n+1}$ may reveal dynamic intent transitions and future-oriented dependencies that cannot be characterized by the single target. Meanwhile, the attempt that generates multiple future items~\cite{GReF} decodes them in parallel from the same context, overlooking the intent transitions that later interactions reveal. To exploit such transition under sequentially dependent supervision, we introduce a \emph{Horizon-aware Preference Generation} (HPG) module, which structurally extends the next-item objective to multiple future horizons.

Specifically, given a behavioral sequence $S_{u}$ of user $u$, we follow the base recommender in Section~\ref{sec:base_recommender} to convert each item $i$ into its aggregated representation $\bm{v}_i$. The original branch (MTP-0) encodes the item-level sequence $\bm{H}^0\!=\![\bm{v}_{1},\bm{v}_{2},\cdots,\bm{v}_{n}]$ as follows:
\begin{equation}
\label{eq:mtp0_encoder}
\bm{Z}^{0}=\mathrm{MTRM}^0(\bm{H}^{0})=[\bm{z}^{0}_{1},\bm{z}^{0}_{2},\cdots,\bm{z}^{0}_{n}],
\end{equation}
where $\mathrm{MTRM}^0(\cdot)$ denotes the multi-layer Transformer backbone within MTP-0. It then feeds the last representation $\bm{z}^{0}_{n}$ into the shared output head $g(\cdot)$ to produce the token-level distributions $\{P^{0}_j\}_{j=1}^{k}$ as in Eq.~\eqref{eq:token_prob} for graph-constrained decoding, where the superscript $b$ in $P^b_j$ indexes the branch hereafter. MTP-0 thus reproduces the base recommender~\cite{RPG}, upon which HPG appends lightweight auxiliary branches as additional supervision pathways while leaving the original architecture and decoding pipeline intact, as detailed below.

Beyond the original branch, HPG appends auxiliary branches to predict items beyond $i_{n+1}$, which not only enriches the supervision on preference transitions but also enables multi-item generation when required. Since both auxiliary branches share the same architecture, we take MTP-1 as an example. Following the sequential MTP design in Eq.~\eqref{eq:mtp_deepseek}, MTP-1 generates the SID of item $i_{n+2}$ by additionally conditioning on the intermediate item $i_{n+1}$, extending the single-step objective to horizon-aware preference modeling while capturing intent transition between adjacent horizons. It fuses the predicted representations $[\bm{z}^{0}_{1},\cdots\!,\!\bm{z}^{0}_{n}]$ from MTP-0 with the embeddings of the shifted sequence $S^1_u\!=\![i_2,\cdots\!,\!i_{n+1}]$ as follows:
\begin{equation}
\label{eq:mtp1_input}
\bm{H}^{1}\!=\!\mathrm{Proj}^{1}\!\left(\![\mathrm{RMSNorm}([\bm{z}^{0}_{1},\!\cdots\!,\!\bm{z}^{0}_{n}\!]);\!\mathrm{RMSNorm}(\![\bm{v}_{2},\!\cdots\!,\!\bm{v}_{n+1}\!])]\right)\!,\!
\end{equation}
where $\mathrm{Proj}^{1}(\cdot)$ projects the concatenated representation into the hidden space of MTP-1. The one-layer Transformer block $\mathrm{TRM}^{1}(\cdot)$ then encodes $\bm{H}^{1}$ into the future-aware representations $\bm{Z}^{1}\!=\!\mathrm{TRM}^{1}(\bm{H}^{1})\!=\![\bm{z}^{1}_{2},\cdots,\bm{z}^{1}_{n+1}]$, whose last entry $\bm{z}^{1}_{n+1}$ is fed into the shared output head $g(\cdot)$ to generate $\{P^1_j\}_{j=1}^k$ for decoding $i_{n+2}$ over the similarity graph $\mathcal{G}$. Likewise, MTP-2 conditions on $\bm{Z}^{1}$ and the shifted sequence $S^2_u\!=\![i_3,\cdots,i_{n+2}]$ to predict $i_{n+3}$, thereby chaining the branches into sequentially dependent horizons. For multi-item generation at inference, the unseen items in each shifted sequence are filled with the top-1 predictions of the preceding branches (\cf Algorithm~\ref{alg:infer}). Sharing the item pool and SID space with MTP-0, both MTP-1 and MTP-2 reuse its token embeddings $\{\bm{E}^{j}\}_{j=1}^{k}$, output head $g(\cdot)$, and decoding graph $\mathcal{G}$, introducing only $\mathrm{Proj}^b(\cdot)$ and $\mathrm{TRM}^b(\cdot)$ per branch to stay lightweight.

\subsection{Verifiable Holistic-Preference Alignment}
\label{sec:VHA}
Although HPG injects future-behavior supervision at the \textit{structural} level, such supervision is distributed across branch-wise objectives and reaches the original branch only indirectly through the auxiliary pathways. Consequently, the decoding representation $\bm{z}^0_n$ that directly generates the recommendation is never explicitly required to internalize the consolidated multi-horizon preference, leaving the future signals under-exploited at the \textit{representation} level. To bridge this gap, we devise a \textit{Verifiable Holistic-Preference Alignment} (VHA) module to consolidate them into a holistic target and explicitly pull $\bm{z}^0_n$ toward it in \textit{representation} space. Specifically, we aggregate the predicted representations of the three branches in Section~\ref{sec:HPG} into a holistic preference representation $\bm{z}_f$ as:
\begin{equation}
    \label{eq:consolidation}
    \bm{z}_f =\mathrm{MLP}_{\text{future}}\big(\big[\mathrm{sg}(\bm{z}^0_n), \mathrm{sg}(\bm{z}^1_{n+1}), \mathrm{sg}(\bm{z}^2_{n+2})\big]\big),
\end{equation}
where $\mathrm{MLP}_{\text{future}}(\cdot)$ is a two-layer projection and $\mathrm{sg}(\cdot)$ denotes the stop-gradient operation, which prevents the alignment objective from back-propagating into the other branches to reshape their supervised signals. Here, the future representations $\bm{z}^1_{n+1}$ and $\bm{z}^2_{n+2}$ convey the evolving intents across horizons, while the present one $\bm{z}^0_n$ serves as an anchor against over-committing to distant and noisier future intents. We then align the decoding representation toward this holistic preference via a representation-level objective:
\begin{equation}
\label{eq:loss_trace}
\mathcal{L}_{\mathrm{trace}} = 1 - \cos\big(\mathrm{Proj}_{0\to f}(\bm{z}^0_n),\, \bm{z}_f\big),
\end{equation}
where $\mathrm{Proj}_{0\to f}(\cdot)$ transports $\bm{z}^0_n$ into the r-dimensional holistic preference space. The stop-gradient in Eq.~\eqref{eq:consolidation} acts only on the construction of $\bm{z}_f$, whereas $\bm{z}_n^0$ in Eq.~\eqref{eq:loss_trace} remains differentiable. Hence, the gradient of $\mathcal{L}_{\text{trace}}$ on the $\bm{z}_f$ side merely calibrates $\mathrm{MLP}_{\text{future}}(\cdot)$, while that on the $\bm{z}_n^0$ side drives the decoding representation to encode holistic preference. Meanwhile, the detached anchor prevents $\bm{z}_f$ from trivially collapsing onto $\bm{z}_n^0$ to shortcut the objective.

However, \textbf{such one-way alignment is inherently \textit{unverifiable}}, as minimizing Eq.~\eqref{eq:loss_trace} is consistent with two opposite outcomes, a faithful one where $\bm{z}^0_n$ internalizes the holistic preference, and a spurious one where $\mathrm{Proj}_{0\to f}(\cdot)$ overfits a shortcut mapping from an uninformed $\bm{z}^0_n$ to $\bm{z}_f$. $\mathcal{L}_{\text{trace}}$ in Eq.~\eqref{eq:loss_trace} alone cannot distinguish between them. Since $\bm{z}^0_n$ and $\bm{z}_f$ reside in different subspaces, this \emph{spurious alignment} is generic rather than incidental: the forward projector $\mathrm{Proj}_{0\to f}(\cdot)$ can simply absorb the discrepancy between the two subspaces within its mapping, satisfying Eq.~\eqref{eq:loss_trace} while leaving $\bm{z}^0_n$ oblivious to the holistic preference (\cf Proposition~\ref{proof:pro}). To suppress such spurious alignment, we require the invertible transport between $\bm{z}^0_n$ and $\bm{z}_f$, which prevents $\mathrm{Proj}_{0\to f}(\cdot)$ from discarding any direction of the holistic space, so that a low alignment loss reflects genuine preference encoding. Therefore, we introduce a backward projector $\mathrm{Proj}_{f\to 0}(\cdot)$ and impose a bidirectional cycle-consistency constraint requiring both round trips to return to their origins: 
\begin{equation}
\label{eq:loss_cyc}
\begin{aligned}
\mathcal{L}_{\mathrm{cyc}} =\ & \big[\,1 - \cos\big(\mathrm{Proj}_{f\to 0}(\mathrm{Proj}_{0\to f}(\mathrm{sg}(\bm{z}^0_n))),\, \mathrm{sg}(\bm{z}^0_n)\big)\,\big] \\
+\ & \big[\,1 - \cos\big(\mathrm{Proj}_{0\to f}(\mathrm{Proj}_{f\to 0}(\mathrm{sg}(\bm{z}_f))),\, \mathrm{sg}(\bm{z}_f)\big)\,\big].
\end{aligned}
\end{equation}
Since both round trips are detached, $\mathcal{L}_{\mathrm{cyc}}$ drives only the projector pair toward a mutually invertible transport between the two subspaces~\cite{Cycle1,Cycle2}. The VHA objective is formulated as $\mathcal{L}_{\mathrm{VHA}} \!=\! \lambda_{\mathrm{trace}}\mathcal{L}_{\mathrm{trace}} \!+\! \lambda_{\mathrm{cyc}}\mathcal{L}_{\mathrm{cyc}}$, where $\lambda_{\mathrm{cyc}}$ balances the alignment objective against round-trip invertibility. Instantiating both projectors as bias-free linear layers, we abbreviate $\bm{z}\!\triangleq\!\bm{z}_n^0$, $P\!\triangleq\!\mathrm{Proj}_{0\to f}(\cdot)$, and $Q\!\triangleq\!\mathrm{Proj}_{f\to 0}(\cdot)$. Since Eq.~\eqref{eq:loss_cyc} is invariant to positive rescalings of $P$ and $Q$, we fix the scale by $\bar{Q}\!\triangleq\!\frac{r}{\mathrm{tr}(PQ)}Q$ whenever $\mathrm{tr}(PQ)\!>\!0$ and define the holistic-side round trip $A\!\triangleq\!P\bar{Q}$, which is invariant to both rescalings. It makes the alignment verifiable, as the invertibility of $A$  is checkable on any trained checkpoint, whereas the one-way objective exposes no such quantity. We characterize what the invertibility guarantees as follows (\cf Appendix for the proof).
\begin{proposition}[Invertible round trip excludes rank-collapse alignment]
\label{proof:pro}
If $A$ is invertible, $\Pi\!\triangleq\!\bar{Q}A^{-1}P$ is a projection onto $\mathrm{range}(\bar{Q})$ along $\ker P$, and every minimizer $\bm{z}_{\min}$ of $\mathcal{L}_{\mathrm{trace}}$ in Eq.~\eqref{eq:loss_trace} satisfies $\Pi\bm{z}_{\min}\!=\!\gamma\,\bar{Q}A^{-1}\bm{z}_f$ for some $\gamma\!>\!0$. Moreover, $\sigma_{\min}\big(P|_{\mathrm{range}(\bar{Q})}\big)\!\geq\!\sigma_{\min}(A)/\|\bar{Q}\|_{\mathrm{op}}\!>\!0$, so $P$ annihilates no direction of $\mathrm{range}(\bar{Q})$.
\end{proposition}
That is, an invertible round trip pins all minimizers of $\mathcal{L}_{\mathrm{trace}}$ to the same projected direction and prevents $P$ from discarding any direction of $\mathrm{range}(\bar{Q})$, thereby excluding the rank-collapse form of spurious alignment. Since $\mathrm{tr}(PQ)\!>\!0$ and $A$ attains full rank on all trained checkpoints, this guarantee holds in practice. In contrast, removing $\mathcal{L}_{\mathrm{cyc}}$ raises $\mathrm{cond}(P)$ by $5.3\!-\!8.4\times$ across datasets, indicating that $P$ absorbs the subspace discrepancy through far more uneven attenuation (\cf Appendix). As Proposition~\ref{proof:pro} constrains only the projected component of $\bm{z}_n^0$, we empirically verify in Section~\ref{subsec:ablation} that $\bm{z}_n^0$ internalizes the holistic preference.

\subsection{Optimization Objective}
\label{sec:objective}
Following the base recommender in Section~\ref{sec:base_recommender}, each branch $b \in \{0, 1, 2\}$ of HPG utilizes its predicted representation $\bm{z}^{b}_{n+b}$ to generate the semantic ID of the target $i_{n+b+1}$ with a token-level MTP loss:
\begin{equation}
\label{eq:HPG_loss}
\mathcal{L}_{b}=-\sum_{j=1}^{k}\log\mathrm{softmax}\left(\bm{E}^{j}\cdot [g(\bm{z}^{b}_{n+b})]_{j}/\tau\right)_{c_{(n+b+1,j)}},
\end{equation}
where $\bm{E}^{j}$ is the embedding matrix of the $j$-th codebook, and $c_{(n+b+1,j)}$ denotes the $j$-th SID of the target item $i_{n+b+1}$ within MTP-$b$. Hence, the overall objective of HPG is $\mathcal{L}_{\mathrm{HPG}} = \mathcal{L}_0 + \lambda_1\mathcal{L}_1 + \lambda_2\mathcal{L}_2$, where $\lambda_1$ and $\lambda_2$ control the contribution of each future-behavior supervision term. Together with $\mathcal{L}_{\mathrm{VHA}}$ in Section~\ref{sec:VHA}, the final optimization objective of \textbf{EchoRec} is formulated as:
\begin{equation}
\label{eq:loss_total}
\mathcal{L} = \mathcal{L}_{\mathrm{HPG}} + \mathcal{L}_{\mathrm{VHA}},
\end{equation}
where the former propagates preferences across horizons and the latter verifiably echoes them back to the decoding representation.\footnote{Due to the space limitation, the detailed training and inference pseudocode and training strategies are provided in Appendix.}

\vspace{0.15cm}
\noindent
\textbf{\emph{Further Discussion.}} Due to space limitations, we provide the computational complexity analysis and discussion of EchoRec's threefold novelty in Appendix.

%% file: Sections/5_Experiments.tex
\begin{table*}[!t]
\centering
\caption{Performance comparison of \textbf{EchoRec} against nine baselines across four datasets. $^{*}$ denotes statistically significant improvements of
\textbf{EchoRec} over its corresponding backbone (\emph{p} $\textless$ 0.01 with paired t-tests).}
\vspace{-0.3cm}
\label{tab:performance_comparison}
\resizebox{\textwidth}{!}{
\begin{tabular}{ccccccccccccccccc}
\toprule
\multirow{2}{*}[-1.5pt]{\textbf{Algorithm}}
& \multicolumn{4}{c}{\textbf{Game}}
& \multicolumn{4}{c}{\textbf{Baby}}
& \multicolumn{4}{c}{\textbf{Arts}}
& \multicolumn{4}{c}{\textbf{PixelRec}} \\
\cmidrule(lr){2-5} \cmidrule(lr){6-9} \cmidrule(lr){10-13} \cmidrule(lr){14-17}
& \textbf{H@10} & \textbf{N@10} & \textbf{H@20} & \textbf{N@20} & \textbf{H@10} & \textbf{N@10} & \textbf{H@20} & \textbf{N@20} & \textbf{H@10} & \textbf{N@10} & \textbf{H@20} & \textbf{N@20} & \textbf{H@10} & \textbf{N@10} & \textbf{H@20} & \textbf{N@20} \\
\midrule
EAGER~\cite{EAGER}
& 0.0385 & 0.0202 & 0.0620 & 0.0261
& 0.0219 & 0.0112 & 0.0382 & 0.0153
& 0.0130 & 0.0093 & 0.0168 & 0.0103
& 0.0257 & 0.0126 & 0.0449 & 0.0174 \\
TIGER~\cite{TIGER}
& 0.0869 & 0.0454 & 0.1335 & 0.0571
& 0.0369 & 0.0194 & 0.0574 & 0.0246
& 0.0617 & 0.0352 & 0.0903 & 0.0424
& 0.0248 & 0.0117 & 0.0457 & 0.0170 \\
LETTER~\cite{LETTER}
& 0.0264 & 0.0125 & 0.0457 & 0.0174
& 0.0214 & 0.0109 & 0.0407 & 0.0157
& 0.0222 & 0.0110 & 0.0353 & 0.0142
& 0.0037 & 0.0019 & 0.0058 & 0.0024 \\
ETEGRec~\cite{ETEGRec}
& 0.0752 & 0.0408 & 0.1142 & 0.0506
& 0.0229 & 0.0116 & 0.0404 & 0.0161
& 0.0555 & 0.0404 & 0.0689 & 0.0438
& 0.0529 & 0.0259 & 0.0972 & 0.0370 \\
GRec~\cite{GRec}
& 0.0766 & 0.0449 & 0.1076 & 0.0527
& 0.0197 & 0.0122 & 0.0262 & 0.0138
& 0.0330 & 0.0200 & 0.0464 & 0.0234
& 0.0053 & 0.0020 & 0.0093 & 0.0030 \\
DimeRec~\cite{GR9}
& 0.0514 & 0.0268 & 0.0817 & 0.0344
& 0.0204 & 0.0104 & 0.0344 & 0.0139
& 0.0347 & 0.0204 & 0.0504 & 0.0244
& 0.0220 & 0.0102 & 0.0444 & 0.0158 \\
GReF~\cite{GReF}
& 0.0817 & 0.0408 & 0.1295 & 0.0529
& 0.0381 & 0.0209 & 0.0524 & 0.0245
& 0.0648 & 0.0372 & 0.0871 & 0.0429
& 0.0263 & 0.0118 & 0.0527 & 0.0185 \\
\midrule
RPG~\cite{RPG}
& 0.0929 & 0.0563 & 0.1250 & 0.0644
& 0.0365 & 0.0245 & 0.0455 & 0.0268
& 0.0625 & 0.0421 & 0.0802 & 0.0466
& 0.0592 & 0.0297 & 0.1052 & 0.0413 \\
\textbf{EchoRec (RPG)}
& \textbf{0.1022*} & \textbf{0.0623*} & \textbf{0.1371*} & \textbf{0.0711*}
& \textbf{0.0425*} & \textbf{0.0295*} & \textbf{0.0525*} & \textbf{0.0320*}
& \textbf{0.0716*} & \textbf{0.0498*} & \textbf{0.0911*} & \textbf{0.0547*}
& \textbf{0.0699*} & \textbf{0.0346*} & \textbf{0.1181*} & \textbf{0.0467*} \\
\textit{Rel. Imp.}
& \textit{+10.01\%} & \textit{+10.66\%} & \textit{+9.68\%} & \textit{+10.40\%}
& \textit{+16.44\%} & \textit{+20.41\%} & \textit{+15.38\%} & \textit{+19.40\%}
& \textit{+14.56\%} & \textit{+18.29\%} & \textit{+13.59\%} & \textit{+17.38\%}
& \textit{+18.07\%} & \textit{+16.50\%} & \textit{+12.26\%} & \textit{+13.08\%} \\
\midrule
SETRec~\cite{SETRec}
& 0.1157 & 0.0610 & 0.1722 & 0.0752
& 0.0322 & 0.0164 & 0.0511 & 0.0212
& 0.0746 & 0.0412 & 0.1086 & 0.0498
& 0.0761 & 0.0371 & 0.1291 & 0.0504 \\
\textbf{EchoRec (SETRec)}
& \textbf{0.1231*} & \textbf{0.0657*} & \textbf{0.1839*} & \textbf{0.0810*}
& \textbf{0.0373*} & \textbf{0.0193*} & \textbf{0.0587*} & \textbf{0.0247*}
& \textbf{0.0838*} & \textbf{0.0468*} & \textbf{0.1197*} & \textbf{0.0559*}
& \textbf{0.0880*} & \textbf{0.0440*} & \textbf{0.1519*} & \textbf{0.0599*} \\
\textit{Rel. Imp.}
& \textit{+6.40\%} & \textit{+7.70\%} & \textit{+6.79\%} & \textit{+7.71\%}
& \textit{+15.84\%} & \textit{+17.68\%} & \textit{+14.87\%} & \textit{+16.51\%}
& \textit{+12.33\%} & \textit{+13.59\%} & \textit{+10.22\%} & \textit{+12.25\%}
& \textit{+15.64\%} & \textit{+18.60\%} & \textit{+17.66\%} & \textit{+18.85\%} \\
\bottomrule
\end{tabular}%
}
\vspace{-0.4cm}
\end{table*}

\section{Experiments}

\subsection{Experimental Setups}

\subsubsection{Dataset} 
To evaluate the effectiveness of \textbf{EchoRec}, we construct four datasets from two platforms, including Game, Baby, and Arts from Amazon Reviews '23 and PixelRec~\cite{PixelRec}. Following traditional GR works~\cite{RPG,GR9}, we chronologically sort each user's historical interactions and extend the leave-one-out evaluation to multi-item prediction task by reserving the last three items for testing, the fourth-to-last item for validation, and the remaining interactions for training. The statistical details are summarized in Table~\ref{tab:dataset}, with construction details in Appendix.

\subsubsection{Baselines.} 
To evaluate the effectiveness of \textbf{EchoRec}, we compare it with nine representative generative recommenders, including \textbf{EAGER}~\cite{EAGER}, \textbf{TIGER}~\cite{TIGER}, \textbf{LETTER}~\cite{LETTER}, \textbf{ETEGRec}~\cite{ETEGRec}, \textbf{GRec}~\cite{GRec}, \textbf{DimeRec}~\cite{GR9}, \textbf{GReF}~\cite{GReF}, \textbf{SETRec}~\cite{SETRec}, and \textbf{RPG}~\cite{RPG}.

\subsubsection{Evaluation Metrics.}
We choose Hit Rate (HR@K) and Normalized Discounted Cumulative Gain (NDCG@K) for evaluation, where k$\in\{10,20\}$. Following conventional GR works~\cite{DiscRec, GR4,GR2}, we adopt the all-ranking paradigm to avoid selection bias. 

\begin{table}[!t]
\centering
\setlength\tabcolsep{5pt}
\caption{Statistics of four real-world datasets.}
\vspace{-0.4cm}
\label{tab:dataset}
\begin{tabular}{c|cccc}
\toprule
\textbf{Dataset} & \textbf{\#User} & \textbf{\#Item}  & \textbf{\#Interactions} & \textbf{Sparsity}  \\ \midrule
Game	& 22,970 	& 30,642 	& 310,554 	& 99.96\% \\
Baby	& 40,501 	& 49,783 	& 485,607 	& 99.98\% \\
Arts	& 20,965 	& 76,985 	& 280,931 	& 99.98\% \\ 
PixelRec & 5,000    & 4,600     & 384,288   & 98.33\% \\
\bottomrule
\end{tabular}
\vspace{-0.5cm}
\end{table}

\subsubsection{Implementation Details}
All methods access the same data during training and testing under identical environments. In training, \textbf{EchoRec} supervises each sample with its next three interactions, where those beyond the training split are padded and masked out, while in testing it uses only MTP-0. Each method is run with five seeds, with details in Appendix.

\subsection{Performance Comparison}
To demonstrate \textbf{EchoRec}'s effectiveness, we implement it with two backbones and compare them with nine baselines across four datasets in Table~\ref{tab:performance_comparison}. We observe that: 
(1) \textbf{EchoRec} significantly outperforms its base recommender on all datasets and metrics, implying that the sequentially dependent supervision of HPG and the verifiable alignment of VHA convert the informative signals within future behaviors into recommendation gains. 
(2) The relative improvements on N@k generally exceed those on HR@k, indicating that \textbf{EchoRec} not only retrieves more correlated target items but ranks them higher, as the holistic preference internalized by VHA refines the decoding representation toward fine-grained preference discrimination.
(3) Among the baselines, the methods with stronger tokenization or decoding mechanisms (\eg SETRec and RPG) generally surpass the earlier GR works, yet their relative strengths fluctuate across datasets, reflecting the sensitivity of the next-item objective to data characteristics. In contrast, \textbf{EchoRec} delivers stable gains over its backbone on every dataset, showing the value of exploiting future behaviors beyond this single target.

\subsection{Ablation Study}
\label{subsec:ablation}
To verify the contribution of each design within our \textbf{EchoRec}, we quantify the contribution of each module in Section~\ref{subsubsec:component_ablation}, and further justify the superiority of the tailored HPG in Section~\ref{subsubsec:multi_horizon}.

\begin{figure}[!t]  
  \centering
  \includegraphics[width=\columnwidth]{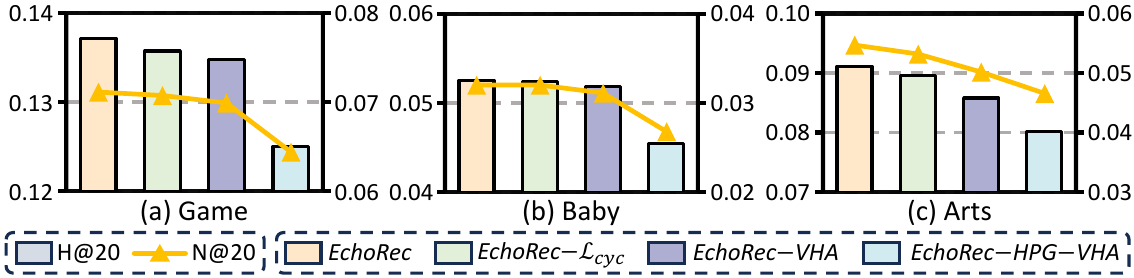}
  \vspace{-0.7cm}
  \caption{Results of EchoRec and its ablation versions on Game, Baby, and Arts. All components are effective.}
\label{fig:Ablation_Study}
\vspace{-0.4cm}
\end{figure}

\subsubsection{Component Ablation}
\label{subsubsec:component_ablation}
To verify the contribution of each module, we compare \textbf{EchoRec} with three variants in Figure~\ref{fig:Ablation_Study}, where \emph{EchoRec$-\mathcal{L}_{cyc}$} keeps the one-way alignment without the cycle-consistency constraint, \emph{EchoRec$-$VHA} retains only the multi-horizon supervision, and \emph{EchoRec$-$HPG$-$VHA} equals the base recommender. We observe that: 
(1) \emph{EchoRec$-$VHA} surpasses \emph{EchoRec$-$HPG$-$VHA} across all datasets, indicating that the sequentially dependent supervision of HPG indeed converts future behaviors into informative signals.
(2) Appending the one-way alignment brings further gains, indicating that \textbf{the consolidated $\bm{z}_f$ carries holistic information beyond the decoding representation rather than trivially replicating $\bm{z}^0_n$}, since aligning toward such a shortcut target would yield no improvement (\cf Appendix).
(3) \textbf{EchoRec} achieves the best performance on all datasets, indicating that the cycle-consistency constraint further suppresses the spurious alignment that the one-way objective alone cannot exclude. 

\begin{figure}[!t]  
  \centering
  \includegraphics[width=\columnwidth]{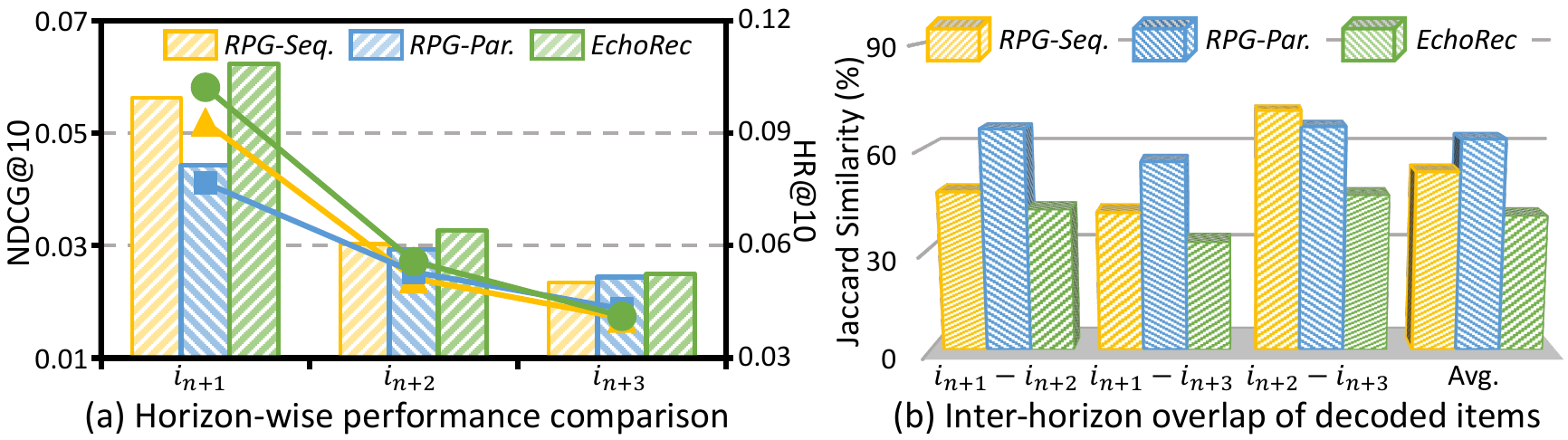}
  \vspace{-0.6cm}
  \caption{Multi-horizon prediction performance of EchoRec against \emph{RPG-Seq.} and \emph{RPG-Par.} on Game, in terms of (a) the horizon-wise performance and (b) the inter-horizon overlap of the decoded items at three future horizons.}
\label{fig:Multi_Horizon}
\vspace{-0.5cm}
\end{figure}

\subsubsection{Multi-Horizon Structure Ablation}
\label{subsubsec:multi_horizon}
To justify the sequential chaining design of HPG, we compare \textbf{EchoRec} with two alternative multi-horizon structures following Section~\ref{subsubsec:preliminary_2}, where \emph{RPG-Par.} decodes all horizons from the shared context and \emph{RPG-Seq.} rolls out the prediction without horizon-specific supervision. As shown in Figure~\ref{fig:Multi_Horizon}, the horizon-wise performance and the inter-horizon overlap examine whether the predictions are accurate and differentiated across horizons, respectively. We notice that: (1) Sequential chaining within HPG improves the future horizons without compromising the immediate one. \textbf{EchoRec} achieves the best performance at $i_{n+1}$ and $i_{n+2}$ on both metrics, and remains competitive at $i_{n+3}$. In contrast, \emph{RPG-Seq.} lags behind across future horizons for lacking any supervision, while \emph{RPG-Par.} suffers from interference at $i_{n+1}$, a degradation that never emerges on \textbf{EchoRec} whose future signals are absorbed without sacrificing the immediate objective. (2) Horizon-specific supervision differentiates the predictions across horizons. \textbf{EchoRec} exhibits the lowest average inter-horizon overlap among all variants, even below the unsupervised rollout of \emph{RPG-Seq.}, indicating that each auxiliary branch captures the intent transition of its own horizon rather than directly replicating the immediate prediction, which resolves the homogenization pitfall identified in Section~\ref{subsubsec:preliminary_2}. These results justify the sequential chaining of our proposed HPG, which additionally endows \textbf{EchoRec} with the multi-item generation ability to forecast a coherent yet non-redundant trajectory within a single model.

\subsection{Robustness Analysis}

\begin{figure}[!t]  
  \centering
  \includegraphics[width=0.98\columnwidth]{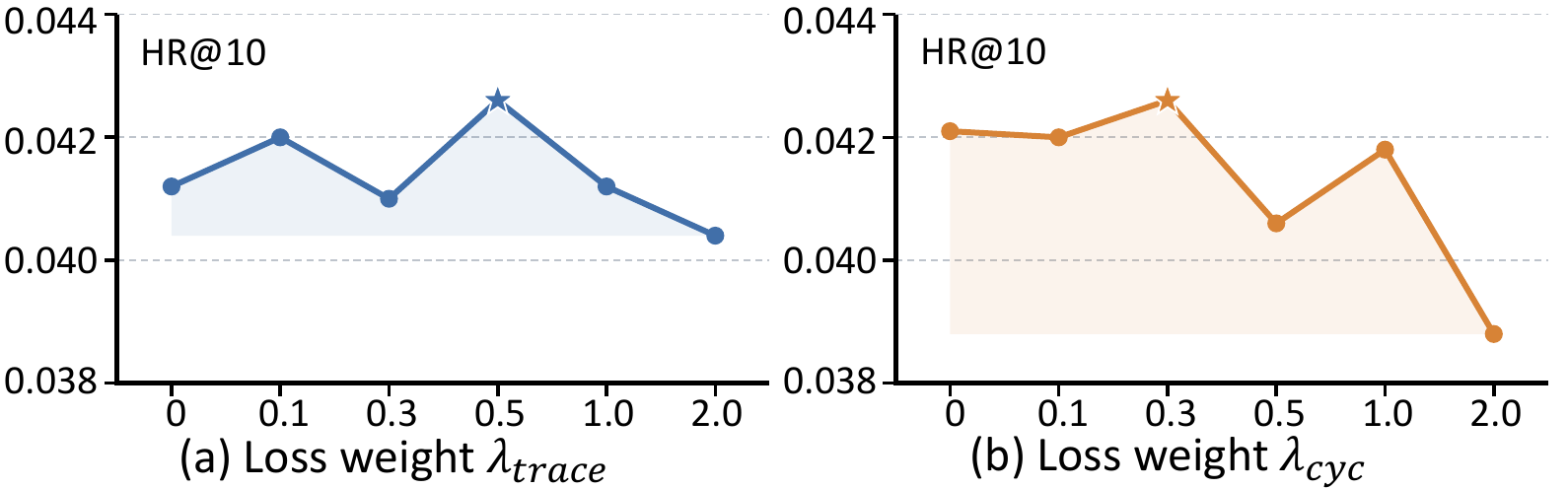}
  \vspace{-0.4cm}
  \caption{Sensitivity of EchoRec to the loss weights (a) $\lambda_{trace}$ and (b) $\lambda_{cyc}$ on Baby, where $\star$ marks the adopted value.}
\label{fig:Parameters}
\vspace{-0.5cm}
\end{figure}

\begin{figure}[!t]  
  \centering
  \includegraphics[width=0.98\columnwidth]{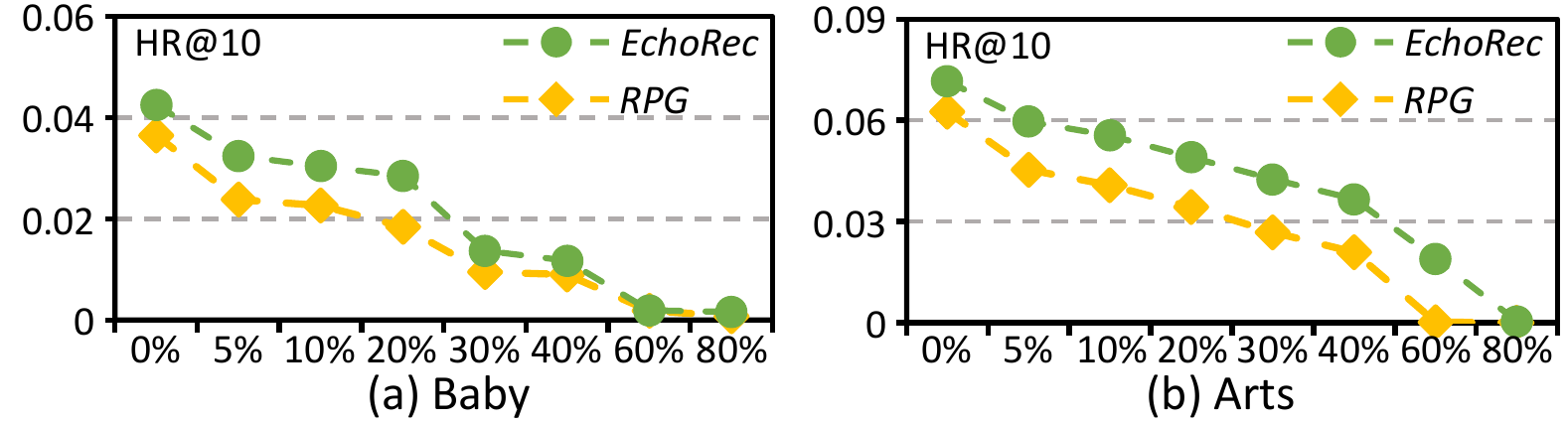}
  \vspace{-0.4cm}
  \caption{Robustness analysis of EchoRec against RPG on Baby and Arts, where the proportion $\eta$ of historical interactions is randomly replaced with sampled items.}
\label{fig:Noisy_Interactions}
\vspace{-0.6cm}
\end{figure}

\subsubsection{Robustness Analysis on Important Parameters}
We examine the sensitivity of \textbf{EchoRec} to the loss weights of VHA on Baby, varying $\lambda_{\text{trace}}$ and $\lambda_{\text{cyc}}$ within $\{0, 0.1, 0.3, 0.5, 1.0, 2.0\}$ in Figure~\ref{fig:Parameters}. Both parameters peak at moderate values ($\lambda_{\text{trace}}\!=\!0.5$ and $\lambda_{\text{cyc}}\!=\!0.3$), as a smaller weight under-exploits the one-way alignment or the cycle constraint, while an over-large one distracts the optimization from the primary next-item objective. \textbf{EchoRec} outperforms its base recommender under all configurations, implying its robustness against the hyper-parameter selection, and we further analyze the initial loss weights of HPG in Appendix.

\subsubsection{Robustness Analysis on Diverse Backbones}
\label{subsubsec:backbone}
Since \textbf{EchoRec} only appends auxiliary branches and optimization objectives to the base recommender without modifying it, it can be seamlessly plugged into diverse backbones. To verify such robustness, we further instantiate it upon SETRec~\cite{SETRec}, whose order-agnostic set identifiers differ fundamentally from RPG. As reported in Table~\ref{tab:performance_comparison}, \textbf{EchoRec (SETRec)} significantly outperforms SETRec across all datasets and metrics, exhibiting the same trend as \textbf{EchoRec (RPG)}. Since the two backbones diverge in both tokenization and decoding strategy, these consistent gains indicate that the multi-horizon supervision of HPG and the verifiable alignment of VHA jointly capture the backbone-agnostic value of future behaviors, rendering \textbf{EchoRec} a plug-and-play framework for GR models.

\subsubsection{Robustness Analysis on Noisy Interactions}
\label{subsubsec:noisy}
To evaluate the robustness of \textbf{EchoRec} against noisy behavioral sequences, we randomly replace a proportion $\eta \!\in\!\{0\%, 5\%, 10\%, 20\%, 30\%, 40\%, 60\%,$ $ 80\%\}$ of items in each sequence with items uniformly sampled from the corpus. As reported in Figure~\ref{fig:Noisy_Interactions}, \textbf{EchoRec} consistently outperforms RPG under mild to moderate noise, and the relative improvement first enlarges with the noise ratio (\eg from 16.4\% at $\eta\!=\!0\%$ to 54.9\% at $\eta\!=\!20\%$) before narrowing under heavier corruption. We attribute this to VHA, which aligns the decoding representation to the consolidated multi-horizon preference rather than individual interactions, so sporadically corrupted items are less likely to distort prediction. 
Only under extreme corruption ($\eta \!\geq\! 60\%$) do both methods degrade to comparable and unusable performance, as the behavioral evidence itself becomes unreliable.

\subsection{In-depth Analysis}
\begin{table}[!t]
\centering
\caption{Efficiency comparison results of \textbf{EchoRec} (RPG) against SETRec and RPG, where ``\#Tra.'' and ``\#Inf.'' denote training and inference, respectively. Latency is averaged per batch, and memory refers to inference GPU memory usage.}
\small
\vspace{-0.3cm}
\label{tab:complexity_analysis}
\resizebox{\linewidth}{!}{
\begin{tabular}{c|c|cccc}
    \toprule
    \multirow{2}{*}[-1.5pt]{\textbf{Dataset}} & \multirow{2}{*}[-1.5pt]{\textbf{Algorithm}}
    & \multicolumn{2}{c}{\textbf{Latency (ms)}}
    & \multirow{2}{*}[-1.5pt]{\begin{tabular}[c]{@{}c@{}}\textbf{Par.}\\\textbf{(M)}\end{tabular}} & \multirow{2}{*}[-1.5pt]{\begin{tabular}[c]{@{}c@{}}\textbf{Mem.}\\\textbf{(MiB)}\end{tabular}}\\ \cmidrule(lr){3-4} 
    & & \textbf{\#Tra.} & \textbf{\#Inf.} &  &  \\
    \midrule
    \multirow{3}{*}{Game}
    & SETRec & 125.61 & 41.57 & 64.15 & 1464.07 \\
    & RPG & 30.11 & 87.44 & 6.00 & 181.04 \\
    & \textbf{EchoRec} & 61.87 & 85.46 & 10.26 & 181.04 \\
    \midrule
    \multirow{3}{*}{Baby}
    & SETRec & 139.81 & 46.46 & 65.37 & 2231.32 \\
    & RPG & 11.25 & 64.13 & 6.00 & 171.54 \\
    & \textbf{EchoRec} & 32.81 & 64.21 & 10.26 & 171.54 \\
    \midrule
    \multirow{3}{*}{Arts}
    & SETRec & 141.68 & 48.33 & 67.11 & 2964.74 \\
    & RPG & 12.50 & 67.32 & 6.00 & 215.40 \\
    & \textbf{EchoRec} & 32.97 & 68.96 & 10.26 & 215.40 \\
    \bottomrule
\end{tabular}}
\vspace{-0.5cm}
\end{table}

\subsubsection{Computational Complexity Analysis}
\label{subsubsec:complexity}
To verify the complexity analysis in Appendix, we compare the per-batch training and inference latency, the number of parameters, and the inference GPU memory usage of \textbf{EchoRec} against SETRec and RPG in Table~\ref{tab:complexity_analysis}. We notice that:
(1) \textbf{The training overhead of EchoRec remains affordable.} Compared with RPG, \textbf{EchoRec} increases the training latency due to the auxiliary branches and alignment objectives, whereas such training-only cost is well amortized as a one-off investment. Its parameters grow from 6.00M to 10.26M, far below the 64.15--67.11M of SETRec, and remain constant across datasets since the auxiliary modules are independent of the numbers of users and items.
(2) \textbf{EchoRec introduces negligible inference overhead.} The inference latency of \textbf{EchoRec} stays nearly identical to RPG on all datasets, empirically confirming that \textbf{all auxiliary components serve as disposable scaffolding discarded at inference}.
(3) \textbf{The memory footprint of EchoRec stays lightweight at serving.} Its inference memory usage remains nearly identical to RPG and substantially below SETRec, as all auxiliary components are discarded and only MTP-0 is activated during serving. Overall, \textbf{EchoRec delivers its gains at marginal training cost while fully preserving the online serving efficiency}.

\begin{figure}[!t]  
  \centering
  \includegraphics[width=0.98\columnwidth]{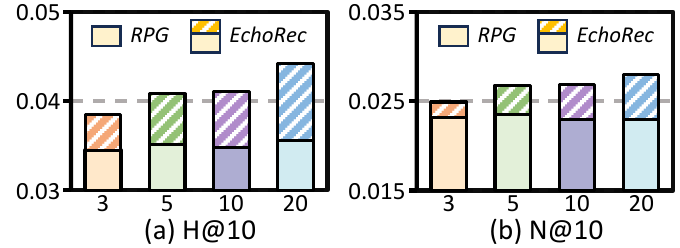}
  \vspace{-0.4cm}
  \caption{Cold-start results on Baby with varying history lengths, where the hatched segments denote EchoRec's gains.}
\label{fig:Cold_Start}
\vspace{-0.5cm}
\end{figure}

\subsubsection{Cold-Start Analysis}
\label{subsubsec:cold_start}
To evaluate whether \textbf{EchoRec} benefits users with scarce histories, we simulate cold-start users on Baby by keeping only the most recent $K\!\in\!\{3, 5, 10, 20\}$ behaviors of each user as the input sequence. As illustrated in Figure~\ref{fig:Cold_Start}, \textbf{EchoRec} consistently outperforms RPG for all $K$, even when only three behaviors are observed. Moreover, the improvement enlarges with $K$ (\eg from 11.6\% at $K\!=\!3$ to 24.2\% at $K\!=\!20$ on H@10), since RPG barely benefits from additional behaviors while \textbf{EchoRec} keeps improving. We attribute this to the specific multi-horizon supervision of our HPG. Since farther horizons depend on long-term preferences beyond the most recent behaviors, learning to predict them encourages \textbf{EchoRec} to exploit longer histories.

\begin{figure}[!t]  
  \centering
  \includegraphics[width=0.98\columnwidth]{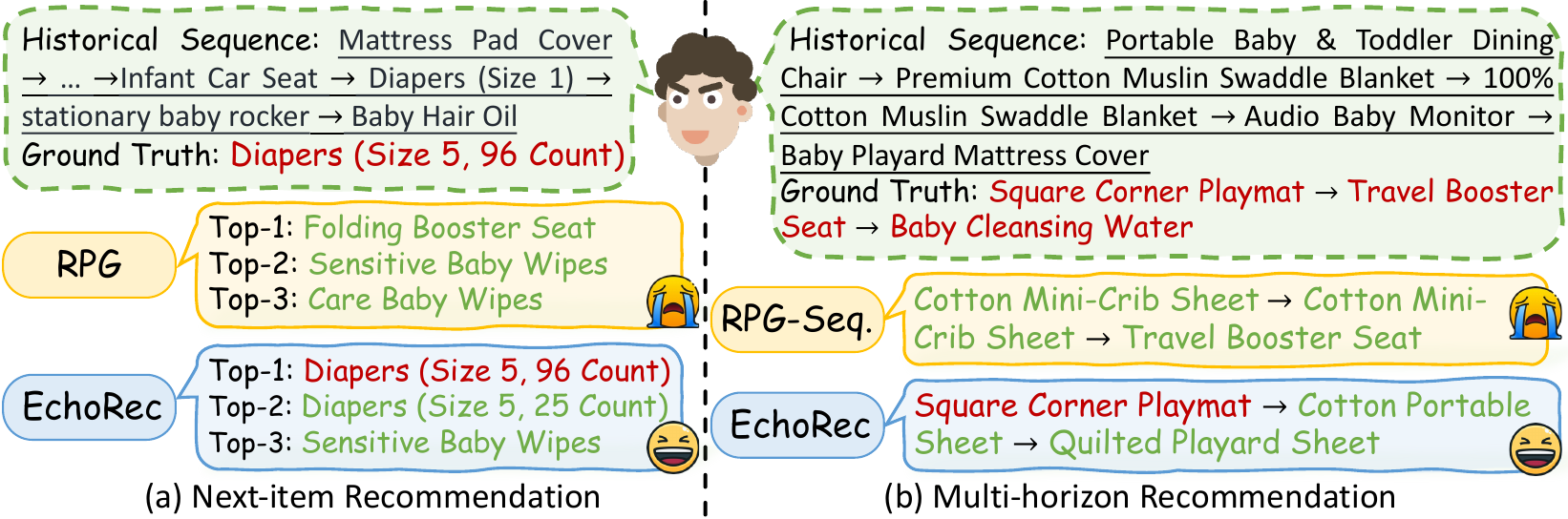}
  \vspace{-0.4cm}
  \caption{Case study of EchoRec on Baby, including (a) the next-item recommendation and (b) the multi-horizon recommendation, where the ground-truth items are marked in red.}
\label{fig:Case_Study}
\vspace{-0.5cm}
\end{figure}

\subsubsection{Case Study}
\label{subsubsec:case_study}
To illustrate how \textbf{EchoRec} benefits from the future-behavior supervision, we show two representative cases from Baby. Figure~\ref{fig:Case_Study} (a) presents a next-item recommendation case, where RPG recommends generally relevant items such as booster seats and baby wipes, whereas \textbf{EchoRec} accurately ranks ``Diapers (Size 5, 96 Count)'' at the top, with a same-series ``Diapers (Size 5, 25 Count)'' immediately after. We attribute this to the holistic preference internalized by VHA, which captures the temporal progression of user demands rather than merely matching the semantics of recent interactions. Figure~\ref{fig:Case_Study} (b) illustrates a multi-horizon recommendation case, where \textbf{EchoRec} correctly generates ``Square Corner Playmat'' at the immediate horizon and continues with diverse yet semantically coherent items, whereas the predictions of \emph{RPG-Seq.} remain close to each other across the first two horizons. These cases qualitatively imply that the multi-horizon supervision of HPG and the verifiable alignment of VHA enable \textbf{EchoRec} to forecast a reasonable trajectory of future behaviors.

%% file: Sections/6_Conclusion.tex
\section{Conclusion}
In this paper, we investigate how to unlock the dense-supervision potential of MTP to enhance generative recommendation, supported by the empirical observation that future behaviors semantically echo the current interaction and thus qualify as informative supervision. Considering that this echo decays under intent transitions and thereby demands sequentially dependent modeling across horizons, we propose \textbf{EchoRec}, which regards preference modeling as an acoustic echoing process. Specifically, HPG converts future behaviors into sequentially dependent supervision through chained MTP branches where each horizon conditions on its predecessor, while VHA consolidates them into the holistic preference and echoes it back through a cycle-consistent projector pair, with theoretical guarantees that exclude the rank-collapse form of spurious alignment. 
Extensive experiments on four datasets demonstrate the effectiveness of our \textbf{EchoRec}, together with its naturally acquired multi-item generation ability. In the future, we plan to explore adaptive horizon selection that calibrates the supervision strength against the decaying semantic echo, and to investigate a mesh-style prediction architecture that captures the fine-grained correlation evolution across both tokens and items.